\documentclass[twocolumn]{aa}
\usepackage{graphicx}
\usepackage{txfonts}

\newcommand{\wfe}{W_{{\rm K}\alpha}}
\newcommand{\efe}{E_{{\rm K}\alpha}}
\newcommand{\sfe}{\sigma_{{\rm K}\alpha}}
\newcommand{\ee}{$e^\pm$}
\newcommand{\g}{$\gamma$}
\newcommand{\lh}{\ell_{\rm h}}
\newcommand{\ls}{\ell_{\rm s}}

\newcommand{\xte}{{\it RXTE}}

\newcommand{\gro}{{\it CGRO}}

\newcommand{\integral}{{\it INTEGRAL}}

\begin{document}

\title{First {\it INTEGRAL\/} observations of Cygnus X-3\thanks{Based on observations with {\it INTEGRAL}, an ESA project with instruments and science data center funded by ESA and member states (especially the PI countries: Denmark, France, Germany, Italy, Switzerland, and Spain), the Czech Republic, and Poland and with the participation of Russia and the US.}}


   \author{O. Vilhu\inst{1,2} \and
          L. Hjalmarsdotter\inst{2} \and A. A. Zdziarski\inst{3}
	  \and A. Paizis\inst{1,4} \and 
	  M. L. McCollough\inst{5}	
          \and V. Beckmann\inst{1,6} \and T. J.-L. Courvoisier\inst{1,7} 
          \and K. Ebisawa\inst{1,8} \and P. Kretschmar\inst{1,9} 
	  \and P. Goldoni\inst{10} \and N. J. Westergaard\inst{11}  
	  \and P. Hakala\inst{2} \and D. Hannikainen\inst{2}}

\offprints{\email{osmi.vilhu@helsinki.fi}}

\institute
{{\it INTEGRAL} Science Data Center, Chemin d'\'Ecogia 16, CH-1290 Versoix, Switzerland
\and 
Observatory, PO Box 14, FIN-00014 University of Helsinki, Finland
\and 
Centrum Astronomiczne im.\ M. Kopernika, Bartycka 18, 00-716 Warszawa, Poland
\and 
CNR-IASF, Sezione di Milano, Via Bassini 15, 20133 Milano, Italy
\and 
Smithsonian Astrophysical Observatory, 60 Garden Street, MS 67, Cambridge, MA 02138-1516, USA
\and
Institut f\"ur Astronomie and Astrophysik, Universit\"at T\"ubingen, Sand 1, 72076 T\"ubingen, Germany
\and 
Geneva Observatory, ch. des Maillettes 51, 1290 Sauverny, Switzerland
\and
Laboratory for High Energy Astrophysics, NASA Goddard Space Flight Center, Greenbelt, MD 20771, USA
\and
Max-Planck-Institut f\"ur Extraterrestrische Physik, Giessenbachstrasse,
  85748 Garching, Germany
\and 
Centre d'Etudes de Saclay, DAPNIA/Service d'Astrophysique, Orme des Merisiers, Gif-sur-Yvette Cedex 91191, France
\and
Danish Space Research Institute, Juliane Maries Vej 30, Copenhagen \O, DK-2100 Denmark}

\date{Received 15 July 2003 /accepted 18 August 2003}

\abstract{We present the first {\it INTEGRAL\/} results on Cyg X-3 from the PV 
phase observations of the Cygnus region. The source was clearly detected by the 
JEM-X, ISGRI and SPI detectors. The {\it INTEGRAL\/} observations were supported 
by simultaneous pointed \xte\/ observations. Their lightcurves folded over the 
4.8 hour binary period are compatible with the mean \xte/ASM and \gro/BATSE 
light curves. We fit our broad-band X-ray/\g-ray spectra with a physical model, 
which represents the first such published model for Cyg X-3. The main physical 
processes in the source are thermal Comptonization and Compton reflection with 
parameters similar to those found for black-hole binaries at high Eddington 
rates.
\keywords{gamma rays: observations -- radiation mechanisms: non-thermal -- stars: individual: Cyg X-3 -- X-rays: binaries -- X-rays: general -- X-rays: stars}}
     
\maketitle

\section{Introduction}

The bright X-ray binary Cyg X-3 was discovered during an early rocket flight already in 1966 (Giacconi et al.\ 1967) but it remains still poorly understood. 
It is a massive system with the donor star and the compact object orbiting each other in a tight orbit. The system is embedded in a dense wind from the donor star, presumably a massive nitrogen-rich Wolf-Rayet star with huge mass loss (van Keerkwijk et al.\ 1992). The nature of the compact object is not known but recent mass estimates suggest it might be a black hole (e.g.\ Schmutz, Geballe \& Schild 1996). 

The system has been observed throughout a wide range of the electromagnetic 
spectrum (e.g.\ McCollough et al.\ 1999). It is one of the brightest Galactic 
X-ray sources, displaying high and low states and rapid variability in X-rays. 
It is also the strongest radio source among X-ray binaries, and shows both huge 
radio outbursts and relativistic jets. The most striking feature in the 
lightcurve is a 4.8-hr quasi-sinusoidal modulation, present both in X-rays and 
infrared. The modulation is believed to reflect the orbital motion of the binary 
with the emission from the X-ray source being scattered by the wind from the 
companion.

\section{Observations and data analysis}

On 2002 Dec.\ 22--23, Cyg X-3 was observed by all the X/\g-ray instruments 
aboard {\it INTEGRAL} -- the JEM-X (Lund et al.\ 2003), IBIS/ISGRI (Lebrun et 
al.\ 2003), IBIS/PICsIT (Di Cocco et al.\ 2003) and SPI (Vedrenne et al.\ 2003). 
The {\it INTEGRAL\/} observations were supported by simultaneous {\it RXTE}/PCA 
and HEXTE observations making possible a comparison of the results from the two 
X-ray telescopes. At the time of the {\it INTEGRAL\/} observations, Cyg X-3 was 
in a relatively high state with the X-ray flux varying between 130 and 330 
mCrab, according to the {\it RXTE}/ASM dwell-by-dwell data. Cyg X-3 was also 
observed in radio by the RATAN and Ryle telescopes. The results of the radio 
observations will be presented elsewhere (Hjalmarsdotter et al., in 
preparation).

\begin{figure}
\centering
 \includegraphics[angle=0,width=8.5cm]{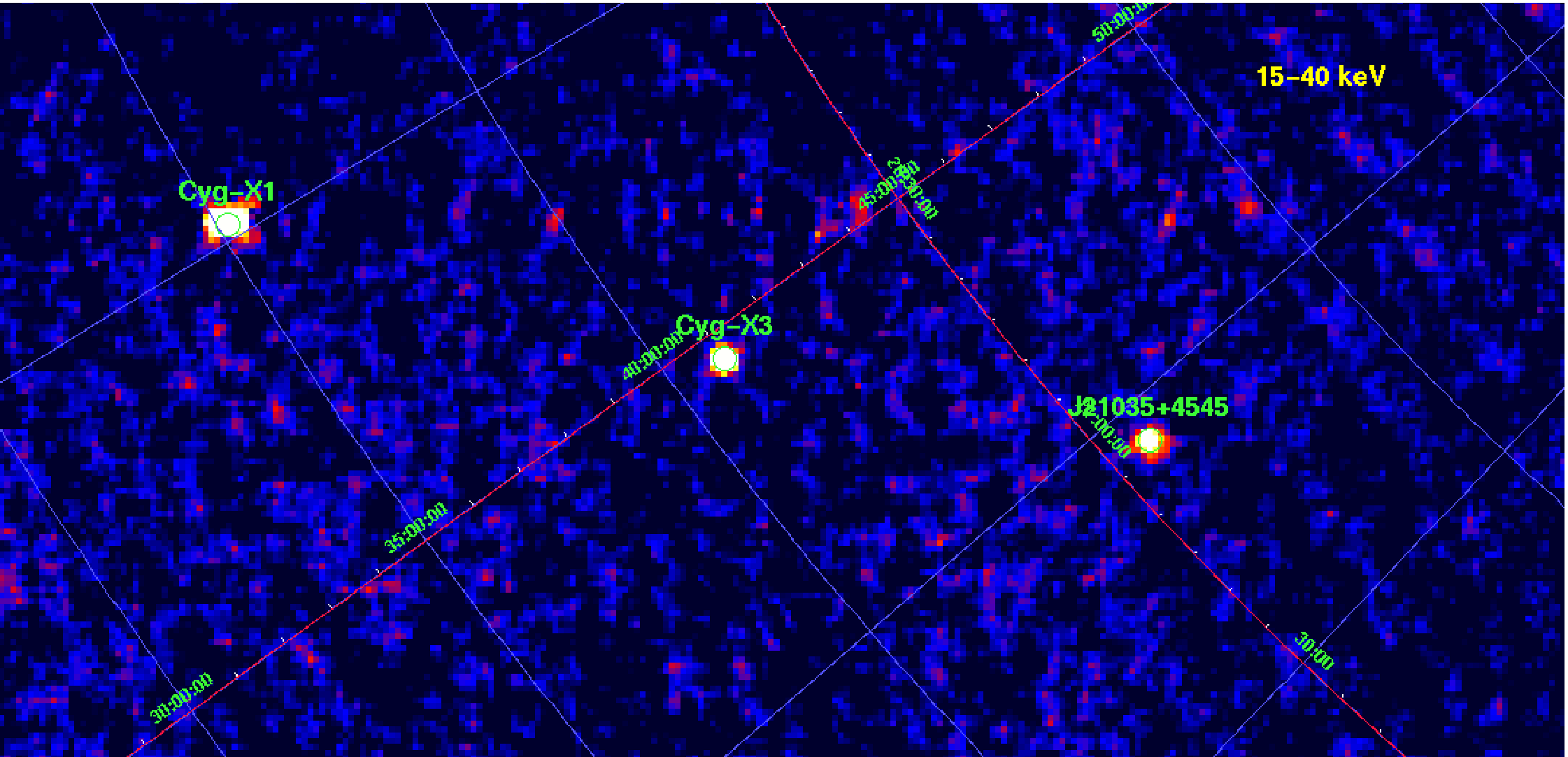} \\  \includegraphics[angle=0,width=8.5cm]{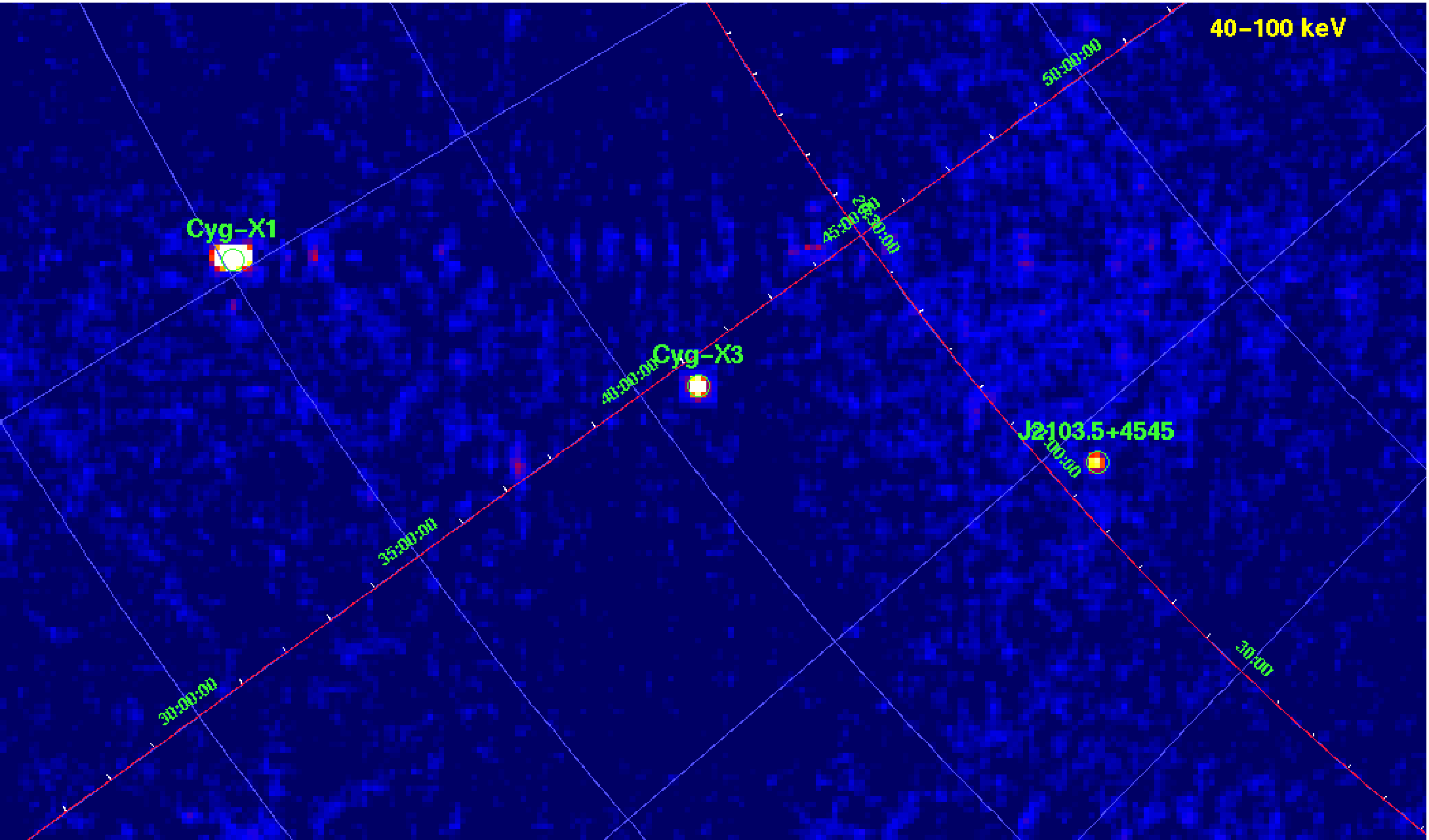} \\
\caption{The ISGRI 20--40 keV (top) and 40--100 keV (bottom) mosaic images showing Cyg X-1 (left), Cyg X-3 (center) and SAX J2103.5+4545 (right).}
\label{isgri}
\end{figure}

\begin{figure}
 \centering
 \includegraphics[angle=270,width=8.5cm]{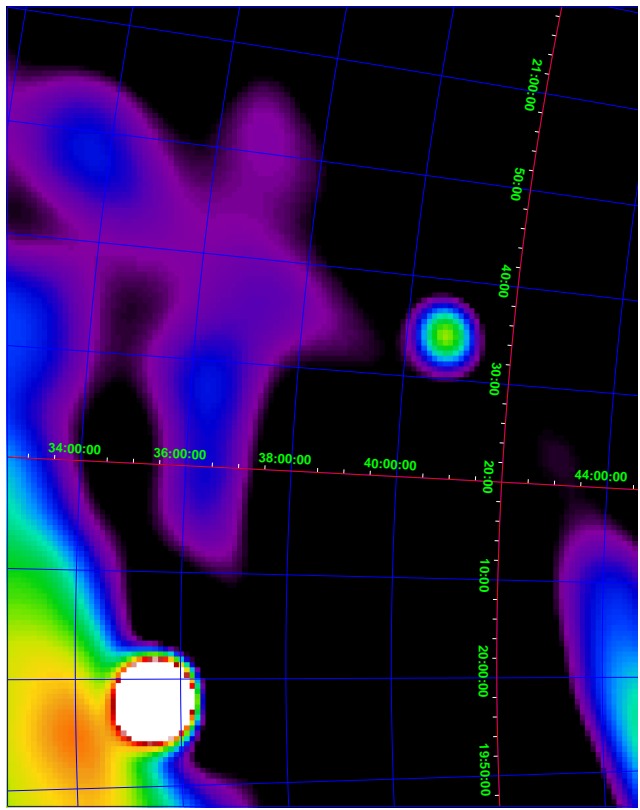}
 \caption{The SPI 34--52 keV image. The brightest source in the lower left corner is Cyg X-1. Cyg X-3 is the source to the upper right.}
 \label{spi}
\end{figure}

\subsection{JEM-X}

The observations were performed with the second (JEM-X2) of the two identical 
X-ray monitors. The data were taken from 43 pointings (science windows), 2200 s 
each, performed between IJD 1085.6--1087.5 (IJD=fractional number of days since 
2000 Jan.\ 1 00.00 UT, which corresponds to MJD = IJD + 51544) during the 23rd 
rev.\ of {\it INTEGRAL\/} (2002 Dec.\ 22--23), which corresponds to 10 
binary periods and the net exposure of $\sim$90 ks. In a half of the pointings, 
Cyg X-3 was in the fully coded field-of-view (FOV, within $2.4\degr$ from  the 
FOV center), while in the rest, it was in the partially coded FOV (within 
$5\degr$ from the FOV center).  In all the pointings, Cyg X-1 and SAX 
J2103.5+4545  (an X-ray pulsar, discovered by Hulleman, in 't Zand \& Heise 1998) were $>\! 10\degr$ from the FOV center and hence the data were not 
contaminated by those sources. The offsets were rather uniformly distributed as 
a function of the binary phase. 

Source spectra were extracted individually per pointing. Then the average 
spectrum was obtained from the sum of the individual spectra weighted by the 
exposure time. The spectral response was Crab-corrected appropriately for this 
time period (instance 0004). The spectra were implicitly background-subtracted 
by a deconvolution algorithm assuming a spatially flat background. We used the 
energy range of 2.6--27 keV for spectral fitting.

\begin{figure*}
 \centering
 \includegraphics[width=14cm]{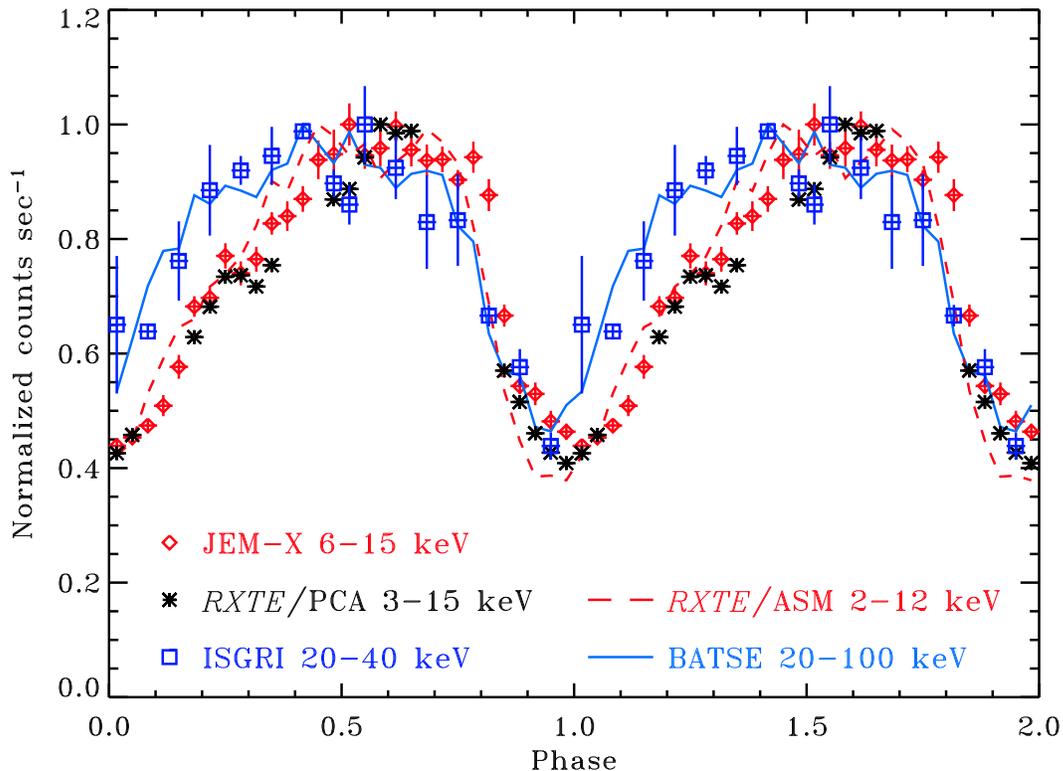}
 \caption{The JEM-X 6--15 keV (red diamonds) and ISGRI 20--40 keV (blue squares) lightcurves folded over the orbital period. We also show the \xte/PCA 3--15 keV data (black asterisks) from simultaneous observations as well as the \xte/ASM 1.5--12 keV (dashed red curve) and \gro/BATSE 20--100 keV (blue curve) phase dependences averaged over several years of monitoring. The count rates are
normalized to the respective maxima.}
\label{lc}
\end{figure*}

\subsection{IBIS/ISGRI}

The ISGRI fully coded FOV is about $9\times 9\degr$, while the partially coded 
FOV extends up to $29\times 29\degr$. Standard spectral extraction is at present 
feasible only in the  fully coded FOV, therefore we limited ourselves to the 
science windows where Cyg X-3 was at a distance of $<\!4.5\degr$ from the 
pointing direction. The selected ISGRI data contain about 40 science windows (of 
an average duration of 2200 s each with exception of two with 500 s duration 
each) for a total duration of $\sim$79 ks.

Imaging analysis was performed using the current version of the Offline 
Scientific Analysis (OSA) software, using the procedure described in Goldwurm et 
al.\ (2003). Cyg X-3 was detected at a high signal-to-noise in the 15--40 and 
40--100 keV energy bands in this as well as in previous ISGRI observations 
(Goldoni et al.\ 2003). The source position was obtained with an offset of 
$<1\arcmin$ with respect to the catalog position. Mosaic images in the 20--40 
and 40--100 keV bands are shown in Fig.\ 1, showing Cyg X-1, Cyg 
X-3, and SAX J2103.5+4545. 

Spectral extraction was performed independently for every science window in 24 
channels linearly rebinned in the 13--200 keV range from a 2048-channel response 
matrix developed at CEA/Saclay. We took the source position as obtained from the 
imaging procedure and then fitted source and background fluxes in each energy 
band. The resulting individual spectra were added to obtain the total spectrum. 
A 5$\%$ systematic error was then added in quadrature to each channel. 

Cyg X-3 was not detected by the IBIS/PICsIT (Foschini, private comm.), which becomes efficient only at 
energies $\ga 250$ keV (Di Cocco et al.\ 2003). Given the low flux from the 
source above 250 keV, a significantly longer exposure would be required for 
detection.

\subsection{SPI}

Out of the 95 dithering pointings taken during the rev.\ 23 on the Cygnus field, 
10 had to be excluded from the SPI analysis as they were either affected by 
strong solar activity or by the radiation belts. This left  85 dithering 
pointings with a total exposure of 169 ks for the present analysis. As the SPI 
data are background-dominated, a careful background substraction is essential in 
order to get reliable results, especially for weak sources. A time-dependent 
background model has been applied to the data, based on the saturated events 
seen by the detector. The image reconstruction was done using the SPI Iterative 
Removal Of Sources program (SPIROS; Skinner \& Connell 2003). To get precise 
flux values, the source positions of the two brightest sources in the field (Cyg 
X-1 and Cyg X-3) have been fixed to their catalogue values. No source confusion 
is expected in the SPI data as there are no other sources visible within 
$3\degr$ around Cyg X-3. The SPI image is shown in Fig.\ 2. 

For spectral extraction, 20 logarithmic bins in the 20--300 keV energy range 
have been used. The instrumental response function used here has 
been derived from on-the-ground calibration and then corrected based on the Crab 
calibration observation. A 5\% systematic error has been added to the spectrum. 

\subsection{PCA/HEXTE}

The {\it RXTE\/} data overlapping with the {\it INTEGRAL\/} observations are 
from two pointings on 2002~Dec.~22--23, of duration of 8224 s (data set 1) and 
9584 s (data sets 2 and 3), respectively. For the second pointing, there was a 
change in the number of PCUs used, which required breaking it into two parts, 
with the lightcurves corrected to the five PCUs. The data sets 1, 2 and 3 then 
cover the binary phases of (0.86--0.06, 0.16--0.34), 0.47--0.67 and 0.82--1.0, 
respectively. Hence, data sets 2 and 3 are from around the maximum and minimum 
phases, respectively. The {\it INTEGRAL\/} spectra, accumulated from all phases, 
should have flux levels in the middle of that from these two sets. 

A 1\% systematic error has been added to the PCA spectra. The relative 
normalization of each of the data sets from the two HEXTE clusters with respect 
to the PCA data has been allowed free in the fits. The response matrices and 
background files have been obtained using the FTOOLS v.\ 5.2.

\section{Lightcurves}

From the JEM-X observations, lightcurves in four energy bands (3--6, 6--10, 
10--15 and 15--35 keV) were created. The ISGRI lightcurve was extracted in the 
20--40 keV band. The lightcurves were folded using the latest published 
ephemeris for Cyg X-3 (Singh et al.\ 2002). 

The results are plotted in Fig.\ 3.  Two middle JEM-X bands were used since 
calibration for the 3--6 and 15--35 keV bands has not been yet consolidated. For 
comparison, we also plot the folded {\it RXTE}/ASM (2--12 keV) and {\it 
CGRO}/BATSE (20--100 keV) lightcurves from monitoring of the source during 
1996--2002 and 1991--2000 respectively. The JEM-X 6--15 keV band shows good 
agreement with the ASM data and the ISGRI 20--40 keV follows the shape of the 
BATSE curve. Modelling of the lightcurves will be presented elsewhere 
(Hjalmarsdotter et al., in preparation). 

\section{Broad-band spectral modelling}

\begin{table*}
      \caption{Model parameters$^{\mathrm{a}}$ for the \xte\/ and \integral\/ spectra. The \xte\/ data sets 1, 2 and 3 are from binary phases (0.86--0.06, 0.16--0.34), 0.47--0.67 and 0.82--1.0, respectively. The {\it INTEGRAL\/} spectra were accumulated from all phases.}
         \label{parameters}
     $$   
         \begin{array}{lccccccccccccc}
            \hline
            \noalign{\smallskip}
{\rm Data} & N_{\rm H,0} & N_{\rm H,1} & f_1 & kT_{\rm s} & \lh/\ls  &  \tau & kT^{\mathrm{b}} & \Omega/ 2\pi & \xi^{\mathrm {c}} & \efe & \wfe & {F_{\rm bol}}^{\rm d} & \chi^2/\nu \\
            \noalign{\smallskip}
& 10^{22}\, {\rm cm}^{-2} & 10^{22}\, {\rm cm}^{-2} && {\rm keV} &&& {\rm keV} &&{\rm erg\, cm}\, {\rm s}^{-1}& {\rm keV}& {\rm eV} & {\rm erg\, cm}^{-2}\, {\rm s}^{-1}\\
            \noalign{\smallskip}
            \hline
            \noalign{\smallskip}
{{\it RXTE}(1)} & 11.8_{-0.8}^{+0.9} & 256^{+9}_{-17} & 0.61_{-0.02}^{+0.02} & 0.37{\rm f} & 0.21_{-0.04}^{+0.02} &  0.23_{-0.02}^{+0.03} & 69 & 1.0_{-0.2}^{+0.2} & 10000_{-5000} & 6.56^{+0.04}_{-0.03} & 400_{-30}^{+30} & 8.2\times 10^{-9} & 239/250\\
             \noalign{\smallskip}
{{\it RXTE}(2)} & 12.8_{-1.0}^{+1.1} & 302^{+20}_{-17} & 0.63_{-0.02}^{+0.02} & 0.41{\rm f} & 0.14_{-0.01}^{+0.02} &  0.16_{-0.02}^{+0.02} & 71 & 1.0_{-0.2}^{+0.2} & 10000_{-5000} & 6.56^{+0.06}_{-0.02} & 310_{-40}^{+30} & 13.5\times 10^{-9} & 250/249\\
             \noalign{\smallskip}
{{\it RXTE}(3)} & 11.1_{-0.3}^{+0.3} & 330^{+13}_{-16} & 0.63_{-0.01}^{+0.01} & 0.45{\rm f} & 0.17_{-0.02}^{+0.02} &  0.19_{-0.02}^{+0.03} & 69 & 1.4_{-0.2}^{+0.3} & 9000^{+1000}_{-3000} & 6.58^{+0.03}_{-0.03} & 510_{-30}^{+30} & 7.0\times 10^{-9} & 174/226\\
             \noalign{\smallskip}
{\it INTEGRAL} & 16.2_{-0.4}^{+0.6} & 334^{+36}_{-37} & 0.44_{-0.02}^{+0.02} & 0.38{\rm f} & 0.18_{-0.03}^{+0.05} &  0.19_{-0.02}^{+0.03} & 75 & 1.1_{-0.2}^{+0.2} & 10000_{-3000} & 6.58^{+0.04}_{-0.04} &230_{-20}^{+20} & 4.3\times 10^{-9} & 291/186\\
             \noalign{\smallskip}
           \hline
         \end{array}
     $$ 
\begin{list}{}{}
\item[$^{\mathrm{a}}$] The uncertainties are for 90\% confidence, i.e., $\Delta \chi^2=2.71$; `f' denotes a parameter fixed in the fit.
\item[$^{\mathrm{b}}$] Calculated from the energy balance, i.e., not a free fit parameter. 
\item[$^{\mathrm{c}}$] Assumed $\leq 10^4$ in the fits.
\item[$^{\mathrm{d}}$] The bolometric flux of the {\it absorbed\/} model spectrum.
\end{list}
\end{table*}

\begin{figure*}
   \centering
   \includegraphics[width=11cm]{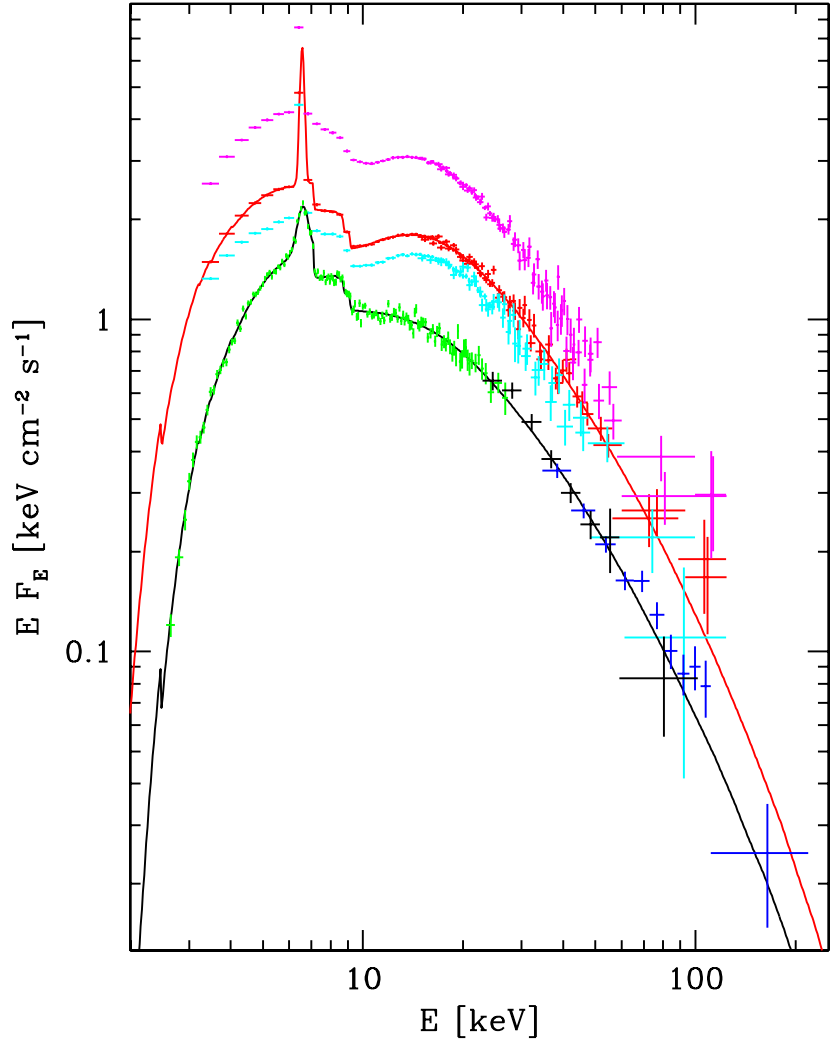}
   \caption{Deconvoled spectra of Cyg X-3. The red, magenta and cyan spectra correspond to the PCA/HEXTE data set 1, 2, 3, respectively. The HEXTE spectra have been renormalized to the flux level implied by the PCA. The model spectrum is shown only for the middle spectrum (data set 1) for clarity. The green, blue and black spectra are from the JEM-X, ISGRI and SPI, respectively, with the model spectrum shown in the black curve. The ISGRI and SPI spectra have been renormalized to the JEM-X data. The intensity levels of the {\it INTEGRAL\/} spectra, which were accumulated over several binary phases, should be close to the mean of the {\it RXTE\/} data sets 2 and 3. The proportions of this figure correspond to equal length per decade on each axis. }
\label{spectra}
    \end{figure*}

The {\it RXTE}/PCA-HEXTE, JEM-X, ISGRI and SPI data were fitted using 
the XSPEC package (Arnaud 1996). We analyze here three \xte\/ spectra and 
the average \integral\/ spectrum. 

We interpret the intrinsic spectra of Cyg X-3 in terms of Comptonization of soft 
X-ray seed photons, assumed here to be a blackbody with a temperature, $T_{\rm 
s}$.  We use a Comptonization model by Coppi (1992, 1999), {\tt eqpair}, 
described in detail by Gierli\'nski et al.\ (1999).  This model was also used to 
fit X-ray spectra of GRS 1915+105 and Cyg X-1 by Vilhu et al.\ (1999), Zdziarski 
et al.\ (2001) and by Poutanen \& Coppi (1998), Gierli\'nski et al.\ (1999), 
Zdziarski et al.\ (2002b), respectively.  In general, the electron distribution 
in this model can be purely thermal or hybrid, i.e., Maxwellian at low energies 
and non-thermal at high energies, if an acceleration process is present.  This 
distribution, including the electron temperature, $T$, is calculated 
self-consistently from the assumed form of the acceleration (if present) and 
from the luminosities corresponding to the plasma heating rate, $L_{\rm h}$, and 
to the seed photons irradiating the cloud, $L_{\rm s}$.  The plasma optical 
depth, $\tau$,  includes a contribution from e$^\pm$ pairs. The importance of 
pairs depends on the ratio of the luminosity to the characteristic size, $r$, 
which is usually expressed in dimensionless form as the compactness parameter, 
$\ell \equiv L\sigma_{\rm T}/(r m_{\rm e} c^3)$, where $\sigma_{\rm T}$ is the 
Thomson cross section and $m_{\rm e}$ is the electron mass.  Hereafter, the 
indices of $\ell$ have the same meaning as those of $L$.

We find all the studied spectra compatible with the hot plasma being completely 
thermal, and with $kT\ll 511$ keV. Then, the \ee\ pair production is negligible, 
and the absolute value of the compactness is only weakly important. Accordingly, 
we assume a constant $\ls=10$ (which is typical for accreting X-ray sources, 
e.g., Gierli\'nski et al.\ 1999).

A complex issue in Cyg X-3 is the structure of its X-ray absorber and the 
presence of additional spectral components in soft X-rays (e.g.\ Molnar \& 
Mauche 1986; Nakamura et al.\ 1993). Given that our data cover energies $\ga\! 
3$ keV only, we neglect any additional soft X-ray components and we use a 
relatively simple model of the absorber. Namely, we assume that an absorbing 
medium with the column density, $N_{\rm H,0}$, fully covers the source, and 
another medium with the column, $N_{\rm H,1}$, covers a fraction, $f_1$, of the 
source. A similar model is often used to model the similarly complex absorption 
of the Seyfert galaxy NGC 4151 (see, e.g.\ Zdziarski et al.\ 2002a). We assume 
the elemental abundances of Anders \& Ebihara (1982). We include Compton 
reflection (Magdziarz \& Zdziarski 1995), parametrized by an effective solid 
angle subtended by the reflector as seen from the hot plasma, $\Omega$, and 
assuming an inclination of $60\degr$. We also include an Fe K$\alpha$ 
fluorescent line, which we model as a Gaussian (with the physical and equivalent 
widths of $\sfe$ and $\wfe$, respectively, and the peak energy at $\efe$). We 
allow the reflecting medium to be ionized, using the ionization calculations of 
Done et al.\ (1992). We define the ionizing parameter as $\xi\equiv 4 \pi F_{\rm 
ion}/n$, where $F_{\rm ion}$ is the ionizing flux and $n$ is the reflector 
density. Given the simplified treatment of the ionized reflection of Done et 
al.\ (1992), we impose a condition of $\xi\leq 10^4$. We assume the temperature 
of the reflecting medium of $10^6$ K.

Given the above approximated treatment of the absorption and ionized reflection, 
the full description of the part  of the spectrum $\la\! 10$ keV is likely to be 
more complex than that given by our model. However, it provides a statistically 
satisfactory description of the data, including the absorbed part of the 
spectrum, and allows us to calculate the broad-band intrinsic spectrum of the 
source. 

During our fits, we have found that the data, covering only photon energies 
$\ga\! 3$ keV, rather poorly constrain the temperature of the seed blackbody 
photons. For example, we get 0.3 keV $\la kT_{\rm s}\la 0.5$ keV within 90\% 
confidence for the \xte\/ data set 1. For simplicity, we fix it at the 
respective best-fit value for each spectrum when determining the confidence 
regions of other parameters. 

On the other hand, our data yield accurate spectral information only for 
energies $\la \! 100$ keV. Thus, they poorly constrain possible electron 
acceleration, which can be present in the plasma in addition to the thermal 
heating. Nonthermal processes are, in fact, clearly observed in some other 
spectral states of Cyg X-3 observed by \xte\/ and the \gro/OSSE (work in 
preparation). However, allowing for the presence of nonthermal electrons 
improves the fit only weakly, e.g.\ by $\Delta \chi^2 \simeq -2$ for  the \xte\/ 
data set 1, and thus it is not required in our models statistically. The 
fraction of the total power supplied to the plasma in electron acceleration is 
constrained to $\la 0.5$, and the power law index, $\Gamma_{\rm acc}$, of the 
acceleration process is not constrained at all at 90\% confidence (typical 
obtained values are $\Gamma_{\rm acc}\sim 2$--4). We also note that the presence 
of nonthermal processes in a very similar state of GRS 1915+105 is required only 
by the data at $\ga\! 100$ keV (see Fig.\ 3a in Zdziarski et al.\ 2001).

The Fe K$\alpha$ line is found to be narrow in all the \xte\/ spectra, with the 
width much below the instrumental resolution of the PCA. The plasma parameters 
obtained are given in Table 1, and the spectra for the 3 data 
sets are shown in Fig.\ 4. 

The \integral\/ spectrum has been found to be rather similar in shape to the 
\xte\/ ones. However, unlike the PCA, the JEM-X data appear to require the Fe 
line to be broadened, with the corresponding decrease of $\Delta \chi^2=-12$. 
Thus, we have decided to allow for the broadening, but, given the limited 
resolution of JEM-X, we kept it then frozen at the best-fit value of $\sfe\simeq 
0.25$ keV. The resulting parameters are given in Table 1, and the spectrum is 
shown in Fig.\ 4. We see that the although the fit parameters are similar to 
those of the \xte\/ fits, the normalization of the JEM-X spectrum is lower than 
that the avarage of the PCA spectra by a factor of $\sim 2$, which is due to instrumental effects, see below. 

\section{Comparison between individual \integral\/ and \xte\/ spectra}

The \integral\/ spectra were accumulated over the entire binary orbit, unlike 
the three {\it RXTE\/} data sets (see Fig.\ 3). The mean flux levels of the {\it 
INTEGRAL\/} spectra should then correspond closely to the mean of the (extreme) 
\xte\/ data sets 2 and 3. However, as we see in Fig.\ 4, the flux level of the 
JEM-X spectrum is about a half of that. Furthermore, the normalizations of the 
HEXTE, ISGRI and SPI differ from each other, with the ISGRI spectrum having the 
flux level about wice of that of the average PCA. Those differences are 
instrumental. To facilitate appropriate corrections to the fluxes from various 
instruments, we list the relative normalizations between the instruments in 
Table 2. The relative ratio between the HEXTE and PCA spectra is consistent with 
previous results. The coefficients involving the \integral\/ data correspond to 
the calibration as of 2003 June. 

\begin{table}
      \caption{The values of the relative normalizations between the flux level from different detectors implied by our data.}
         \label{norms}
     $$   
         \begin{array}{lcc}
            \hline
            \noalign{\smallskip}
{\rm Instruments} & E_1\!-\!\!E_2\, [{\rm keV}]^{\mathrm{a}} & {\rm ratio}^{\mathrm{b}}  \\
            \noalign{\smallskip}
          \hline
            \noalign{\smallskip}
{\rm HEXTE\,0/PCA(2)} & 18\!-\!\!120 & 0.71\pm 0.04\\
{\rm HEXTE\,1/PCA(2)} & 18\!-\!\!120 & 0.70\pm 0.05\\
{\rm JEM\!-\!\!X/PCA(2)}^{\mathrm{c}} & 2.6\!-\!\!27 & 0.33\pm 0.01\\
{\rm JEM\!-\!\!X/PCA(3)}^{\mathrm{c}} & 2.6\!-\!\!27 & 0.65\pm 0.01\\
{\rm ISGRI/JEM\!-\!\!X} & 35\!-\!\!220 & 4.7\pm 0.2\\
{\rm SPI/JEM\!-\!\!X} & 23\!-\!\!100 & 2.3\pm 0.1\\
            \noalign{\smallskip}
\hline
         \end{array}
     $$ 
\begin{list}{}{}
\item[$^{\mathrm{a}}$] The energy range of the first of the two compared instruments used to derive the ratio.
\item[$^{\mathrm{b}}$] The uncertainties are 1-$\sigma$.
\item[$^{\mathrm{c}}$] The JEM-X spectrum should have about the same level as an average of the PCA (2) and (3) spectra, which implies the renormalization factor of $\sim$0.44. 
\end{list}
\end{table}

The slopes of various spectra agree with each other well, as shown in Fig.\ 4. 
The ISGRI and SPI data fitted by a power law over the energy ranges given in 
Table 2 yield  the photon index of $\Gamma=3.6 \pm 0.1$ with $\chi^2/\nu=34/22$, 
$9/9$, respectively. On the other hand, those data show hardenings at lower 
energies (not shown in Fig.\ 4), which appear related to residual inaccuracies 
of the present response matrices. 

\section{Conclusions}

Cyg X-3 was clearly detected by all three X/\g-ray instruments on board {\it 
INTEGRAL}. The JEM-X 6--15 keV lightcurve folded over the orbital period shows 
good agreement with the {\it RXTE}/PCA 3--15 keV and {\it RXTE}/ASM 1.5--12 keV 
lightcurves. The ISGRI 20--40 keV folded lightcurve matches the {\it CGRO}/BATSE 
20--100 keV one. A difference in the light curve profile between energies above 
and below $\sim$15 keV is indicated. 

For the first time, we fit Cyg X-3 X/\g-ray spectra with a physical model. The 
main radiative processes implied by the \integral\/ and \xte\/ data are thermal 
Comptonization and Compton reflection. The obtained intrinsic spectrum appears similar to that of GRS 1915+105 at a similar Eddington ratio. 

At the time of writing, there are apparent differences in the normalizations 
between the {\it RXTE}/PCA, JEM-X, SPI and ISGRI spectra. Calibrations and 
responses at this stage are constantly being improved. Cyg X-3 will be further 
observed with {\it INTEGRAL\/} as a part of the Core  as well as the Guest 
Observer programmes.

\begin{acknowledgements}
Authors from the Observatory of the University of Helsinki acknowledge
the Academy of Finland, TEKES, and the Finnish space research programme
ANTARES for financial support in this research. AAZ has been supported by KBN grants 5P03D00821, 2P03C00619p1,2, PBZ-KBN-054/P03/2001, and the Foundation for Polish Science. We would like to thank M. Revnivtsev for the tools to produce the ISGRI mosaics and J. Poutanen (the referee) for valuable comments. We acknowledge quick-look results provided by the {\it RXTE}/ASM team.
\end{acknowledgements}

\end{document}